\documentclass[11pt]{article}
\usepackage{amssymb, amsmath}
\usepackage{color, pdfcolmk}
\usepackage{graphics}
\usepackage{graphicx}
\usepackage{amssymb}
\usepackage{epsf}
\usepackage{verbatim}

\newtheorem{theorem}{Theorem}

\newtheorem{definition}[theorem]{Definition}
\newtheorem{lemma}[theorem]{Lemma}

\newtheorem{corollary}[theorem]{Corollary}
\newtheorem{example}[theorem]{Example}

\newcommand{\eps}{\varepsilon}

\begin{document}

\title{Large attractors in cooperative bi-quadratic Boolean networks. Part II.}

\author{
 \and
Winfried
Just\footnote{Department of Mathematics, Ohio University.}
\ and
German A. Enciso\footnote{Mathematical Biosciences Institute, Ohio State University, and Harvard Medical School, Department of Systems Biology.
\newline
This material is based upon work supported by the National Science Foundation
under Agreement No. 0112050 and by The Ohio State University.}
}

\maketitle

\begin{abstract}
Boolean networks have been the object of much attention, especially since S. Kauffman proposed them in the 1960's as models for gene regulatory networks.   These systems are characterized by being defined on a Boolean state space and by simultaneous updating at discrete time steps. Of particular importance for biological applications are networks in which the indegree for each variable is bounded by a fixed constant, as was stressed by Kauffman in his original papers.

An important question is which conditions on the network topology can rule out
 exponentially long periodic orbits in the system.
In this paper we consider cooperative systems, i.e. systems with positive feedback interconnections among all variables, which in a continuous setting guarantees a very stable dynamics.  In Part~I of this paper we presented a construction that shows that for an arbitrary constant $0<c<2$ and sufficiently large $n$ there exist $n$-dimensional Boolean cooperative networks in which both the indegree and outdegree of each for each variable is bounded by two (bi-quadratic networks) and which nevertheless contain periodic orbits of length at least $c^n$.

In this part, we prove an inverse result showing that for sufficiently large $n$ and for $0<c<2$ sufficiently close to 2, any $n$-dimensional cooperative, bi-quadratic Boolean network with a cycle of length at least $c^n$ must have a large proportion of variables with indegree 1.  Such systems therefore share a structural similarity to the systems constructed in Part~I.
\end{abstract}




\section{Introduction}

In this note we continue our study of the existence of exponentially long periodic orbits in bi-quadratic cooperative Boolean systems.  The motivation for our interest in this problem was described in Part~I of this paper \cite{PartI}.

For a positive integer $n$, let $[n]:= \{1, \ldots , n\}$.
An $n$-dimensional \emph{Boolean dynamical system} or \emph{Boolean network} is a pair $(\Pi, g)$, where $\Pi = \{0, 1\}^{[n]}$ and $g: \Pi \rightarrow
\Pi$.  A state $s(t)$ at time $t$ will be denoted by $s(t) = [s_1(t), \ldots , s_n(t)]$, or simply
 $s = [s_1, \ldots , s_n]$ if time-dependency is ignored.  We will have

\begin{equation}\label{discrete}
s(t+1) = g(s(t)).
\end{equation}

 The \emph{cooperative order } on $\Pi$ is the partial order relation defined by $s \leq r$ iff
 $s_i \leq r_i$ for all $i \in [n]$.  The system is \emph{cooperative} if $s(t) \leq r(t)$ implies
 $s(t+1) \leq r(t+1)$.

 We associate a directed graph $D$ with vertex set $[n]$ with the system.  A pair $<i, j>$ is in the arc set of $D$
 iff there exist states $s, r \in \Pi$ such that $s_i < r_i$ and $s_k = r_k$ for all $k \neq i$ with the property that
 $(g(s_i))_j < (g(r_i))_j$.  We will say that the system is \emph{quadratic} if  the indegree  of all vertices
 in $D$ is at most two.  We call the system \emph{$b$-quadratic} if it is quadratic and  the outdegree of all vertices
 in $D$ is at most $b$, where $b$ is a positive integer. A $2$-quadratic system is called  \emph{bi-quadratic.}  More generally, for positive integers $b, r$ we define a $(b,r)$-Boolean system as a system in which the indegree of all vertices
 in $D$ is bounded by~$r$ and the outdegree of all vertices in $D$ is bounded by $b$.

In Part~I of this paper \cite{PartI} it was shown that for every $0 < c < 2$ and sufficiently large $n$ there exist $n$-dimensional bi-quadratic cooperative Boolean networks that contain periodic orbits of length at least $c^n$. These systems were constructed by conceptualizing a small subset $M$ of
the variables as a Turing machine and the set $T$ of the remaining $n - |M|$ variables as $L$ circular tapes in such a way that $M$ writes successive codes of the integers $0, \ldots , \lfloor c^n\rfloor$ on the tape.  A judicious choice of coding allowed us to find examples where the whole system is cooperative and bi-quadratic.

While the metaphor of a Turing machine acting on one or several tapes is certainly appealing to the human mind, it is an intriguing question whether
our construction is, in some sense, \emph{the only way} of producing cooperative bi-quadratic Boolean systems with very long cycles.  Here we show that this is indeed the case.

Consider an $n$-dimensional Boolean system $(\Pi, g)$.  Then $g = [g_1, \ldots , g_n]$.
Taking our motivation from Boolean models of gene regulatory networks, we call $g_k$ \emph{the $k$-th regulatory function.}
If $(\Pi, g)$ is quadratic, then $g_i$ depends on at most two variables $i_k, j_k$.  If, in addition, $(\Pi, g)$ is cooperative, and if $i_k \neq j_k$, then we must have
$g_k = s_{i_k} \wedge s_{j_k}$ or $g_k = s_{i_k} \vee s_{j_k}$. In these two cases,  we will say that \emph{$g_k$ is strictly quadratic.}  The only other possibility
is $g_k = s_{i_k}$; we will say in this case that \emph{$g_k$ is monic.}
Note that if $g_k$ is constant then we get identical dynamics along attractors if we replace it with the monic function $g_k = s_k$.
Since transient states are irrelevant for our results, we will wlog assume that the indegree of each variable is at least one.
In a more general setting, we will say that $g_k$ is monic iff the indegree of $k$ in the digraph $D$ associated with the system is one.

Now consider a sequence of variables $k_1, \ldots , k_m$ such that $g_{k_{i+1}} = s_{k_i}$ for all $i \in [m-1]$.  The dynamics of the system on these
variables is analogous to that of a memory tape that advances by one position at each time step.  A new value may be written to position $k_1$ at each time step, and this value may be read $\ell$ time steps later by some regulatory function off position $k_{\ell + 1}$.  If $k_m = k_1$, the tape is
`read-only,' and a constant regulatory function can be considered a special case of a `read-only' tape of length~one.
Also, a tape could split into two or more branches (say, if $g_2= s_1$, \ $g_3= g_4 = s_2$, $g_5= s_3$, $g_6 = s_4$, etc.,
then a branching would occur at the second variable),
but the values on these branches would eventually be only copies of each other.
Thus any cooperative system that contains monic regulatory functions can be conceptualized as a Turing machine
acting one or more tapes, possibly branching or of varying lengths.  This observation motivates the following definition.

\begin{definition}\label{turingdef}
We call an $n$-dimensional Boolean system  an \emph{$(M, n)$-Turing system} if at least $n - M$ of the regulatory functions are monic.
\end{definition}

While every $n$-dimensional Boolean system is an $(n,n)$-Turing system in the sense of the above definition, we will use this expression to
highlight the fact that if $M < n$, the roles of the `machine' and the `tapes' can be neatly separated.
Note that we do not require in Definition~\ref{turingdef} that the system be cooperative.  Admittedly, if also monic regulatory functions
$g_k = \neg s_{i_k}$ may occur in the system, then the connection with the Turing machine metaphor becomes more tenuous, but we will still
use this terminology as a convenient way to formulate our results.

As indicated above, the systems constructed in \cite{PartI} are $(M(n),n)$-Turing systems such that $\lim_{n \rightarrow \infty} \frac{|M(n)|}{n} = 0$. The main result we will prove here shows that if $c$ is sufficiently close to $2$, then in \emph{every} bi-quadratic cooperative Boolean system with a periodic orbit of length at least $c^n$ the `tapes' must contain the vast majority of all variables.

\begin{theorem}\label{Turingthm}
Let $\alpha > 0$ and let $b$ be a positive integer. Then there exists a positive constant $c < 2$ such that for sufficiently large $n$, every $n$-dimensional $b$-quadratic cooperative Boolean system with a periodic orbit of length at least $c^n$ is an $(\alpha n, n)$-Turing system.
\end{theorem}

We will actually prove a more general result than Theorem~\ref{Turingthm}.
The \emph{bias}  $\Lambda$  of a
Boolean function is the fraction of input vectors for which the function outputs 1.  Note that the strictly quadratic cooperative Boolean functions $x \wedge y$ and $x \vee y$ have bias $\Lambda = 0.25$ and $\Lambda = 0.75$ respectively, whereas monic Boolean functions have bias $\Lambda = 0.5$.
There are two strictly quadratic Boolean functions with bias $\Lambda = 0.5$, namely the exclusive-or function and the equivalence function, but these are not cooperative and cannot occur in cooperative Boolean systems.  On the other hand, there are cooperative Boolean functions with bias $\Lambda = 0.5$ that depend on three input variables; an important example is the function that takes the value $1$ iff the majority of input variables have the value $1$.

We will say that a Boolean system $(\Pi, g)$ is \emph{$\eps$-biased} if every non-monic regulatory function has bias $\Lambda$ with
  $|\Lambda - 0.5| \geq \eps$.  It follows from the above discussion that cooperative quadratic Boolean systems are $0.25$-biased, but in general,
  cooperative Boolean systems with regulatory functions that can take three or more inputs need not be $\eps$-biased for any
  $\eps > 0$.  We will prove the following generalization of Theorem~\ref{Turingthm}.

\begin{theorem}\label{GenTthm}
Let $\varepsilon, \alpha > 0$ and let $b, r$ be positive integers. Then there exists a positive constant $c(\eps, \alpha, b, r) < 2$ such that for every $c > c(\eps, \alpha, b, r)$ and sufficiently large $n$,
 every $n$-dimensional $\eps$-biased $(b,r)$-Boolean system with a periodic orbit of length at least $c^n$ is an $(\alpha n, n)$-Turing system.
\end{theorem}

A Boolean function $\varphi$ that depends on variables $x_1, \ldots , x_\ell$ is
\emph{canalyzing}
if there exists one input variable $x_c$, called the \emph{canalyzing variable,} a Boolean value $u$ called the \emph{canalyzing value,}
and a Boolean value $v$ called the \emph{canalyzed value} such that
$\varphi(x_1, \ldots , x_\ell) = v$ whenever $x_c = u$.  Clearly, all three kinds of regulatory functions that are permitted in quadratic cooperative
Boolean systems are canalyzing.
Another important example of a quadratic canalyzing Boolean function is the implication $x \rightarrow y$.  It is easy to see that a canalyzing Boolean function has bias $\Lambda = 0.5$ iff it is monic.  Since there are only finitely many
Boolean functions on any fixed number of inputs, the following is an immediate consequence of Theorem~\ref{GenTthm}.

\begin{corollary}\label{CanTthm}
Let $\alpha > 0$ and let $b, r$ be positive integers. Then there exists a positive constant $c < 2$ such that for sufficiently large $n$,
 every $n$-dimensional  $(b, r)$-Boolean system with a periodic orbit of length at least $c^n$, and in which all regulatory functions are canalyzing, is an $(\alpha n, n)$-Turing system.
\end{corollary}

\section{Proof of Theorem~\ref{GenTthm}}

We will prove Theorem~\ref{GenTthm} in two stages.  In the first stage of the proof we will show that very large subsets of the state space $\Pi$ of an $n$-dimensional Boolean system must be \emph{balanced} in a sense that will be defined shortly.  In the second stage of the proof we will show that if $S$ is the set of states in a periodic orbit of an $\eps$-biased $(b,r)$-Boolean system and $S$ is sufficiently balanced, then only a small fraction of the regulatory functions can be non-monic.

\subsection{Balanced subsets of the state space}

Let $\Pi = \{0,1\}^{[n]}$ be the state space of an $n$-dimensional Boolean system.    Let $S = \{s^\ell: \, \ell \in L\}$ be a sequence of (not necessarily pairwise distinct) elements of $\Pi$.   If the elements of $S$ happen to be pairwise distinct, then we will speak of $S$ being a
\emph{subset} of $\Pi$.

To illustrate the key idea of this section, let $i \in [n]$ and consider the ratio

$$\zeta_i (S) = \frac{|\{\ell \in L: \, s^\ell_i = 1\}|}{|L|}.$$

If $\beta, \gamma > 0$, then we will say that $S$ is \emph{$\beta$-$\gamma$-$1$-balanced} if $|\{i \in [n]: \ |\zeta_i(S) - 0.5| \geq \gamma\}| < \beta n$.

More generally, let $r \in [n]$ and $\sigma : [r] \rightarrow \{0,1\}$.  For $r$-element subsets $I = \{i_1, \ldots , i_r\}$ of $[n]$ with
$i_1 < \dots < i_r$
we define ratios
$\xi_{I}^{\sigma}(S)$ as follows:

$$\xi^{\sigma}_{I} (S) = \frac{|\{\ell \in L: \, \forall u \in [r] \ s_{i_u}^\ell = \sigma(u)\}|}{|L|}.$$

Define

$$\zeta^*_I(S) = \max \{2^{-|I|} - \xi^{\sigma}_{I}(S): \sigma \in  \{0,1\}^{[r]} \}.$$

If $\beta, \gamma > 0$, then we will say that $S$ is \emph{$\beta$-$\gamma$-$r$-balanced} if for every family $P$ of pairwise disjoint
subsets $I$ of $[n]$ with $|\bigcup P| \geq \beta n$ and $1 \leq |I| \leq r$ for each $I \in P$ there exists $I \in P$ such that
$\zeta^*_I(S) < \gamma$.

We will prove the following.

\begin{lemma}\label{balancedlemma}
Let $r$ be a positive integer, $\beta, \gamma > 0$ and assume $\gamma < 2^{-r}$. Let

$$\lambda(\gamma, r) = \left(\frac{1- 2^{-r}}{1- 2^{-r}+\gamma}\right)^{1- 2^{-r}+\gamma}\left(\frac{2^{-r}}{2^{-r}-\gamma}\right)^{2^{-r}-\gamma},$$
and let $c$ be a constant such that
$$c > 2(\lambda(\gamma, r))^\beta.$$
Then  for sufficiently large $n$, every subset $S$ of $\{0,1\}^{[n]}$ of size $\geq c^n$ is  $\beta$-$\gamma$-$r$-balanced.
\end{lemma}

\noindent
\textbf{Proof:}  Let $\beta, \gamma, r$ be as in the assumptions, and assume throughout this argument that~$n$ is a sufficiently large positive
integer.  Let $\varrho > 0$, let $1 < c < 2$, and let $\delta$ be such that
$1 + \varrho\gamma < \delta < 1 + 2\varrho\gamma$ and $\delta c^n$ is an integer.
Let us assume that $S = \{s^\ell: \, \ell \in [\delta c^n]\}$ is a sequence of randomly and independently (with replacement) chosen states in
 $\{0,1\}^{[n]}$ of length
$\delta c^n$.
We will treat  $\xi_I^\sigma$ and $\zeta^*_I$ as random variables and temporarily suppress their dependence on $S$ in our notation.

Let $v \in [r]$. For fixed $I = \{i_1, \ldots , i_v\}$ with $i_1 < \dots < i_v$ and $\sigma \in \{0, 1\}^{[v]}$ we define

$$\eta_I^\sigma = \frac{\sum_{\ell=1}^{\delta c^n} \eta_{I\ell}^\sigma}{\delta c^n},$$

where $\eta_{I\ell}^\sigma = 0$ if $ s^\ell(i_u) = \sigma (u)$ for all $u \in [v]$, and $\eta_{I\ell}^\sigma = 1$ otherwise.

Clearly, the mean value of $\eta^\sigma_{I}$
is $E(\eta^\sigma_{I}) = 1 - 2^{-v}$.  Note that $2^{|I|} - \xi^\sigma_I \leq \eps$ iff $\eta^\sigma_{I} -E(\eta^\sigma_{I}) \geq \eps$, and hence
$\zeta^*_I \geq \eps$ iff $\eta^\sigma_{I} -E(\eta^\sigma_{I}) \geq \eps$ for at least one
$\sigma \in \{0,1\}^{[v]}$.

We want to estimate $Pr(\eta^\sigma_{I} -E(\eta^\sigma_{I}) \geq \eps)$ for any given fixed $\eps > 0$.
Note that the random variables $\eta_{I\ell}^{\sigma}$ take values in the interval $[0,1]$ and are independent.  This allows us to use the
following inequality of \cite{Hoeffding} (see also \cite{Okamoto, Chernoff} for the special case we are considering here).

\begin{lemma}\label{Jansoneq}
Let $X_1, X_2, \dots, X_m$ be independent random variables such that $0 \leq X_i \leq 1$ for $i \in [m]$ and let
$X = (X_1 + \dots + X_m)/m$.  Let $\mu = E(X)$ and let $0< \eps < 1-\mu$.   Then

\begin{equation}\label{Hoeffeqn}
Pr(X - \mu \geq \eps) \leq \left(\left(\frac{\mu}{\mu+\eps}\right)^{\mu+\eps}\left(\frac{1-\mu}{1 - \mu-\eps}\right)^{1 - \mu-\eps}\right)^m
\leq e^{-2\eps^2m}.
\end{equation}
\end{lemma}

We will assume until further notice that $\eps < 2^{-\nu}$ and thus satisfies the assumptions of~(\ref{Hoeffeqn}).
Both bounds in~(\ref{Hoeffeqn}) are of the form $\lambda^m$ for some $0 < \lambda \leq e^{-2\eps^2} < 1$.  For the moment, assume that
$\lambda$ is such such a constant, and let
$m = \delta c^n$.  Now it follows from~(\ref{Hoeffeqn}) that

$$Pr(\eta_{I}^\sigma - 1 + 2^{-\nu} \geq \eps) \leq \lambda^{\delta c^n}.$$

This implies the following estimate for $\zeta^*_I$:

$$Pr(\zeta^*_I  \geq \eps) \leq 2^v \lambda^{\delta c^n}.$$

Now fix $k < n$ and consider $k$ pairwise disjoint subsets $I_1, \ldots , I_k$ of cardinality $\leq r$ each.
The random variables $\zeta^*_{I_1}, \ldots , \zeta^*_{I_k}$ are independent.  It follows that

$$Pr(\forall m \in [k] \ \zeta^*_{I_m} \geq \eps) \leq 2^{rk} \lambda^{k\delta c^n}.$$

Let $k = \beta n$ and let $A$ be the event that there exists family $P$ of pairwise disjoint
subsets $I$ of $[n]$ with $|\bigcup P| \geq \beta n$ and $1 \leq |I| \leq r$ for each $I \in P$ such that $\zeta^*_{I} \geq \eps$ for each
 $I \in P$.  The number of eligible families $P$ is bounded from above by $\binom{n}{r}^{\beta n} < n^{r\beta n}$.
 Thus the probability of the event $A$ can be estimated as

$$Pr(A) < (2n)^{r\beta n}\lambda^{\beta n \delta c^n}.$$

Now note that by Stirling's formula the number of subsets of $\Pi$ of size $c^n$ satisfies

$$\binom{2^n}{c^n} < \frac{2^{nc^n}}{c^n!} < \frac{1}{2}\frac{2^{nc^n} e^{c^n}}{c^{nc^n}} = \frac{1}{2}\left(\frac{2e^{\frac{1}{n}}}{c}\right)^{nc^n}.$$

Moreover, note that

$$\lim_{n \rightarrow \infty} (2n)^{\frac{r\beta}{c^n}} = 1.$$

Thus for

\begin{equation}\label{cestimate}
c > 2\lambda^{\beta\delta}
\end{equation}

and $n$ sufficiently
large, we will have

$$(2n)^{\frac{r\beta}{ c^n}}\lambda^{\beta\delta} < \left(\frac{2e^{\frac{1}{n}}}{c}\right)^{-1}.$$

This in turn implies that for sufficiently large $n$ and $c$ as in~(\ref{cestimate})

\begin{equation}\label{punch1}
Pr(A) < (2n)^{r\beta n}\lambda^{\beta n\delta c^n} =  \left((2n)^{\frac{r\beta}{c^n}}\lambda^{\beta \delta}\right)^{n c^n}
< (\frac{2e^{\frac{1}{n}}}{c})^{-n  c^n} < \frac{1}{2\binom{2^n}{c^n}}.
\end{equation}

Now let us fix $\eps$ such that $0 < \eps < \gamma$.  Since $\gamma < 2^{-r}$, the assumptions of Lemma~\ref{Jansoneq}
will be satisfied for this choice of $\eps$. Let $B = B(S)$ be the set of the first $c^n$ pairwise distinct elements of the sequence $S$, if in fact
$S$ has at least $c^n$ pairwise distinct elements, and let $B$ be undefined otherwise.  Let us make a few observations:

\begin{enumerate}
\item Let $N = \{\ell \in [\delta c^n]: \ \exists 1 \leq j < \ell \ s^j = s^\ell\}$ be the number of entries in $S$ that duplicate
a previous entry.  Note that $B$ is defined iff $N \leq (\delta - 1) c^n$.  In particular, by the choice of $\delta$, the set
$B$ is defined as long as  $N \leq \varrho\gamma c^n$.

\item Note that the expected value of $N$ can be estimated, for sufficiently large $n$, fixed $c < 2$, and $0 < \varrho < \frac{2-c}{2c\gamma}$, as

$$E(N) \leq \sum_{\ell \in [\delta c^n]} \frac{\ell-1}{2^{n}} < \frac{\delta^2 c^{2n}}{2^n} = o(1) c^n.$$

In particular, $E(N) < \frac{\delta - 1}{2}c^n$.

\item Now it follows from Markov's Inequality

$$\frac{\delta - 1}{2}c^n > E(N) \geq Pr(N > (\delta -1)c^n)(\delta-1)c^n$$
 that for fixed $c$ and sufficiently large $n$, the set $B$ will be defined with probability
$> 0.5$.

\item Assume $B$ is defined.  Observe that for  each subset $I$ of $[n]$ and
$\sigma \in \{0, 1\}^{[|I|]}$ we have

\begin{equation}\label{BSeqn}
\frac{\eta^\sigma_I(B)}{\delta} \leq \eta^\sigma_I(S)\leq \frac{\eta^\sigma_I(B) + \delta - 1}{\delta}.
\end{equation}

The first inequality in~(\ref{BSeqn}) turns into equality if $\eta_{I\ell}^\sigma = 0$ whenever $s^\ell$ is outside of $B$;
the second inequality in~(\ref{BSeqn}) turns into equality if $\eta_{I\ell}^\sigma = 1$ whenever $s^\ell$ is outside of $B$.
It follows from the relationship between the $\eta^\sigma_I(S)$'s and $\zeta^*_I(S)$ that
$$ \frac{\zeta^*_I(B)}{\delta} \leq \zeta^*_I(S)\leq \frac{\zeta^*_I(B) + \delta - 1}{\delta}.$$
By choosing $\varrho$ sufficiently close to $0$, we can choose $\delta$ as close to one as we need, and
our choice of $\eps < \gamma$  implies that for $\delta$ sufficiently close to one the inequality $\zeta^*_I(S) < \eps$ will
 imply the inequality $\zeta^*_I(B) < \gamma$.
\end{enumerate}

But if there is \emph{any} subset $B$ of size $c^n$ of $\Pi$ that is not $\beta$-$\gamma$-$r$ balanced, then
 this subset will be exactly as likely to be equal to $B(S)$ as any
other subset of $\Pi$ of the same size. By point 3 above, the probability that $B(S)$ exists is greater than $0.5$, and thus the probability that
$B(S)$ exists and is equal to $B$ must be
at least $0.5 \binom{2^n}{c^n}^{-1}$. 
But point~4 above implies that if $B$ is not $\beta$-$\gamma$-$r$ balanced, then $B(C) = B$ implies that the event $A$
has occurred, with contradicts inequality~(\ref{punch1}).

We derived the contradiction under the assumption that $c$ satisfies inequality~(\ref{cestimate}).  Now assume
$$c > 2(\lambda(\gamma, r))^\beta$$ as in the assumption of the lemma.  Then we can choose
$\eps$ sufficiently close to $\gamma$ and $\lambda = \lambda(\eps, r)$
so that inequality~(\ref{cestimate}) will hold as well for any $\delta > 1$. By choosing
$\delta$ sufficiently close to one we will get a contradiction whenever $B$ exists and satisfies $\zeta^*_I(B) \geq  \gamma$.
This proves Lemma~\ref{balancedlemma}. $\Box$

\subsection{Systems with balanced periodic orbits}

\begin{lemma}\label{orbitlem}
Let $b,r$ be positive integers, let $0 <  \eps, \tau < 0.5$, let $(\Pi, g)$ be an $n$-dimensional Boolean system, and let $S$ be a periodic orbit of
$(\Pi, g)$.
Let $k \in [n]$ be such that the bias~$\Lambda$ of $g_k$ satisfies $|\Lambda - 0.5| \geq \eps$, and let $I$ be the set of input variables
of $g_k$.  Then either $\zeta^*_I(S) \geq \frac{\tau}{2^{|I|}}$ or $\zeta^*_{\{k\}}(S) \geq (1-\tau)\eps - \frac{\tau}{2}$.
\end{lemma}

\noindent
\textbf{Proof:} Assume wlog that $\Lambda \geq 0.5 + \eps$; the proof in the case when $\Lambda \leq 0.5 - \eps$ is symmetric.
Suppose that $\zeta^*_I < \frac{\tau}{2^{|I|}}$.  Then there exists a subset $S^* \subseteq S$ with $|S^*| \geq (1 - \tau)|S|$ such that $\eta_I^\sigma(S^*) = 2^{-|I|}$ for each $\sigma \in \{0,1\}^I$.
We conclude that
\begin{equation*}
\begin{split}
&|S|\zeta_k = |\{s \in S: \ s_k = 1\}| = |\{s \in S: \ g_k(s) = 1\}| \geq |\{s \in S^*: \ g_k(s) = 1\}|\\
&\geq (1- \tau)|S|\Lambda \geq \left(1- \tau\right)|S|\left(0.5 + \eps\right) > |S|\left(0.5 + (1-\tau)\eps - \frac{\tau}{2}\right),
\end{split}
\end{equation*}
and the inequality  $\zeta^*_{\{k\}} \geq (1-\tau)\eps - \frac{\tau}{2}$ follows. $\Box$

\begin{lemma}\label{cyclelemma}
Let  $(\Pi, g)$ be an $n$-dimensional $\eps$-biased $(n,r)$-Boolean system, let $0 < \tau < \frac{\eps}{\eps + 0.5}$, let $\gamma = \frac{\tau}{2^{r}}$, $\gamma^* = (1-\tau)\eps - \frac{\tau}{2}$, and let $\beta, \beta^* > 0$.  Assume $S$ is the set of states in a periodic orbit of $(\Pi, g)$ so that $S$ is both $\beta$-$\gamma$-$r$-balanced and
$\beta^*$-$\gamma^*$-$1$-balanced.  Then  there exists a subset $J \subseteq [n]$ of size $|J| < (\beta  + r \beta^*) n$ with the property that every non-monic regulatory function $g_k$ has at least one input variable in $J$.
\end{lemma}

\noindent
\textbf{Proof:} Let $K = \{ k \in [n]: \ \zeta^*_{\{k\}}(S) \geq \gamma^*\}$. The assumption on $S$ implies that $|K| < \beta^* n$.

Let $J_0$ be the set of inputs of the variables in $K$.  Then $|J_0| < r\beta^* n$.

Let $K^+ = [n] \backslash K$ and let $\{k_1, \ldots , k_p\} \subseteq K^+$ be a set of variables maximal with respect to the property that
$g_{k_q}$ is non-monic for every $q \in [p]$ and the sets $I_q$ of inputs of $g_{k_q}$ are pairwise disjoint.  Let $J_1 = \bigcup_{q \in [p]} I_q$.

By Lemma~\ref{orbitlem} and the choice of $K^+$, for each $q \in [p]$ we must have $\zeta^*_{I_q}(S) \geq \frac{\tau}{2^{r}}$.  Thus the assumption on $S$ implies that $|J_1| < \beta n$.

On the other hand, by maximality of $\{k_1, \ldots , k_p\}$, every non-monic regulatory function~$g_k$ must have at least one input in the set
$J:= J_0 \cup J_1$, and the lemma follows. $\Box$

\bigskip

Now let $(\Pi,g), \eps, \alpha, b, r$ be as in the assumptions of Theorem~\ref{GenTthm}, let $\gamma, \gamma^*$ be as in the assumptions of Lemma~\ref{cyclelemma}, and assume that $\beta, \beta^* > 0$ satisfy

\begin{equation}\label{beteqn}
\beta + r\beta^* = \frac{\alpha}{b}.
\end{equation}

Let $\lambda(\gamma, r), \lambda(\gamma^*, 1)$ be as in Lemma~\ref{balancedlemma}.  Then we will have

\begin{equation}\label{ceps}
c(\eps, \alpha, b, r) \leq \max\{2(\lambda(\gamma, r))^{\beta}, 2(\lambda(\gamma^*, 1))^{\beta^*}\}.
\end{equation}

To see this, assume $n$ is sufficiently large and $S$ is a periodic orbit of $(\Pi,g)$ of length at least $c^n$, where
$c$ exceeds the right-hand side of~(\ref{ceps}).  Then Lemma~\ref{balancedlemma} implies that
$S$ is $\beta$-$\gamma$-$r$-balanced and $\beta^*$-$\gamma^*$-$1$-balanced 
and thus satisfies the assumptions of Lemma~\ref{cyclelemma}.  Let~$J$ be as in the conclusion of
Lemma~\ref{cyclelemma}.  Note that at most $b|J| < \alpha n$ regulatory functions can have inputs in $J$, and it follows that $(\Pi, g)$ is an
$(\alpha n, n)$-Turing system.

Note that by the second inequality in~(\ref{Hoeffeqn}) we will in particular have

\begin{equation}\label{cepsw}
c(\eps, \alpha, b, r) \leq \max\{2e^{-2\gamma^2\beta}, 2e^{-2(\gamma^*)^2\beta^*}\}.
\end{equation}

This concludes the proof of Theorem~\ref{GenTthm}. $\Box$

\bigskip

We formulated  Theorem~\ref{GenTthm} as a qualitative result about existence of a constant and wrote the proof so as to make the argument as transparent as possible.
In the remainder of this paper we will use the notation $c(\eps, \alpha, b,r)$ as shorthand for the largest real number for which the conclusion of Theorem~\ref{GenTthm} holds.

To arrive at more precise estimates of
$c(\eps, \alpha, b, r)$, we defined $\gamma = \frac{\tau}{2^r}$, $\gamma^* = (1-\tau)\eps - \frac{\tau}{2}$, and
wrote a simple MatLab program for numerically exploring the values of the right-hand side of~(\ref{ceps}) for
$\tau \in (0, \frac{\eps}{\eps+0.5})$ and $\beta^* \in (0, \frac{\alpha}{br})$. Note that the value of $\beta$ is not a free parameter as it is
given by~(\ref{beteqn}).

For the case of bi-quadratic cooperative systems, when $\eps = 0.25$ and $b = r = 2$,
we found an almost perfect linear approximation of the right-hand side of~(\ref{ceps}).

\begin{equation}\label{cepslin}
c(0.25, \alpha, 2, 2) \leq \max\{2(\lambda(\gamma, r))^{\beta}, 2(\lambda(\gamma^*, 1))^{\beta^*}\} \approx 2 - 0.0041\alpha.
\end{equation}

\section{Three additional observations}

Theorem~\ref{Turingthm} cannot be proved without the assumption of some bound on the outdegrees in the associated digraph:  note that in the following example the system in question is not necessarily bi-quadratic.

\begin{example}\label{quadrex}
Let $0 < c < 2$. Then for all sufficiently large $n$ there exists an $n$-dimensional quadratic cooperative Boolean system $(\Pi, g)$ that has only strictly quadratic regulatory functions and contains a periodic orbit of length $c^n$.
\end{example}

\noindent
\textbf{Proof:} Let $(\Sigma, f)$ be a bi-quadratic cooperative Boolean system  of dimension $n-2$ that contains a periodic orbit of length $c^n$, as constructed in \cite{PartI}.  Let $\Pi = \{0, 1\}^{[n]}$, let $g_k = f_k$ whenever $k < n-1$ and $f_k$ is strictly quadratic, let
$g_k = s_{i_k} \wedge s_n$ whenever $k < n$ and $f_k = s_{i_k}$, and let $g_{n-1} = g_n = s_{n-1} \wedge s_n$.  Then $(\Pi, g)$ is cooperative, quadratic, and has only strictly quadratic regulatory functions.

Now let $s \in \Sigma$
be a state in a periodic orbit of length at least $c^n$ of $(\Sigma , f)$, and define a state  $s^* \in \Pi$ by
$s^* = [s_1, \ldots, s_{n-2}, 1, 1]$.  Then the orbit of $s^*$ in $(\Pi, g)$ has the same length as the orbit of
$s$ in $(\Sigma, f)$. $\Box$

\begin{example}\label{c=sqrt2}
Let $0 < c < \sqrt{2}$. Then for all sufficiently large $n$ there exists an $n$-dimensional bi-quadratic cooperative Boolean system $(\Pi, g)$ that has only strictly quadratic regulatory functions and contains a periodic orbit of length $c^n$.
\end{example}

\noindent
\textbf{Proof:} If $c$ is as in the assumption, then $c^2 < 2$.  Let $n$ be sufficiently large and wlog even, and let $(\Sigma, f)$ be a bi-quadratic cooperative Boolean system  of dimension $\frac{n}{2}$ that contains a periodic orbit of length $c^{2n}$, as constructed in \cite{PartI}.  Define an $n$-dimensional
Boolean system $(\Pi, g)$ in such a way that for every monic regulatory function $f_k = s_{i_k}$ we have
$g_k = g_{n/2 +k} = s_{i_k} \wedge s_{i_{n/2 + k}}$.  This can be done in such a way that $(\Pi, g)$ remains bi-quadratic, cooperative, and no new monic regulatory functions are introduced, while the dynamics of the system on $\left[\frac{n}{2}\right]$ coincides with the dynamics of $(\Sigma, f)$ for all initial states with $s_{i}(0) = s_{n/2 + i}(0)$ for all $i \in \left[\frac{n}{2}\right]$. $\Box$
\bigskip

Finally, recall that a subset $T = \{k_1, \ldots , k_m\}$ of $[n]$ with $g_{k_{i+1}} = s_{k_i}$ for all $i \in [m-1]$ and
$g_{k_1} = s_{k_m}$ of a Boolean system $(\Pi, g)$ can be considered a `read-only' tape.  Let us define a \emph{generalized read-only tape} as
a set of variables
$T = \{k_1, \ldots , k_m\} \subseteq [n]$ such that $k  \in T$, for all $i \in [m-1]$ we have
$g_{k_{i+1}} = s_{k_i}$ or $g_{k_{i+1}} = \neg s_{k_i}$,
and
$g_{k_1} = s_{k_m}$ or
$g_{k_1} = \neg s_{k_m}$. Let us define the \emph{read-only part} or \emph{strictly monic part} $R(g)$ of a
Boolean system $(\Pi, g)$ as the union of all its generalized read-only tapes.  The following theorem shows that Boolean systems with large read-only parts cannot have very long periodic orbits.

\begin{theorem}\label{readonlythm}
Let $\delta > 0$ and let $c > 2^{1-\delta}$.  Then for sufficiently large $n$, no $n$-dimensional Boolean system $(\Pi, g)$ with $|R(g)| \geq \delta n$ can have a periodic orbit of length $\geq c^n$.
\end{theorem}

\noindent
\textbf{Proof:} First note that if $T = \{k_1, \ldots , k_m\}$ is a generalized tape of length $m$, then the dynamics of the system on $T$ is can be described by cyclical shifts, possibly with negations in some positions.  Thus given any initial state $s(0)$ of the system, the vector
$[s_{k_1}(t), \ldots , s_{k_m}(t)]$ can take  at most $2m$ distinct values throughout the trajectory of $s(0)$.
Since $R(g) = \{\ell_1, \ldots , \ell_{R}\}$ is the union of pairwise disjoint generalized tapes $T_1, \ldots , T_v$ with
$|T_1| + \dots +|T_v| = |R(g)|$, it also follows from the same observation that
 the vector $[s_{\ell_1}(t), \ldots , s_{\ell_{R}}(t)]$ can take  at most $2lcm \left(\{|T_1|, \ldots , |T_v|\}\right)$ distinct values throughout the trajectory of $s(0)$, where $lcm$ stands for the least common multiple.

If $P(N)$ denotes the maximum value of $lcm\left( \{m_1, \ldots , m_r\}\right)$ with $\sum_{i=1}^r m_i = N$,
then
$P(N)= e^{ \sqrt{N \ln N} (1+o(1)) }$  as $N\to \infty$ (see Chapter 13 of \cite{Landau}). It follows that for any given initial state $s(0)$, the vector of values of the variables in $R(g)$ can take at most $2e^{ \sqrt{|R(g)| \ln |R(g)|} (1+o(1)) }$ different values in any periodic orbit.
 Thus under the assumptions of the theorem, the number of different states in any orbit is bounded from above by $2^{(1 - \delta)n}2e^{\sqrt{n \ln n}(1 + o(1)}$, which is less than $c^n$ for $c$ as in the assumption and sufficiently large~$n$. $\Box$

\section{Discussion}

The dependence of the length of periodic orbits, or limit cycles,
 in Boolean networks on the network architecture has been the subject of
several empirical studies.  In these studies, 
random networks are drawn from the ensemble of all Boolean networks with prescribed restrictions on the architecture, and the dynamics is studied by simulating the trajectories for a sample of initial states. Very long limit cycles are a hallmark of the \emph{chaotic regime,} whereas relatively short limit cycles are a hallmark of the \emph{ordered regime.}

It has been observed that the dynamics tends to become more ordered and less chaotic if the number of inputs for each regulatory functions is small
\cite{origins}, if the regulatory functions are strongly biased \cite{DerridaStauffer, WeisbuchStauffer}, if all regulatory functions are (nested) canalyzing functions \cite{nestcan}, or if there are few negative feedback loops \cite{Sontag:Laubenbacher}.

Our approach is different from that of the empirical studies cited above in that we try to find conditions on the network architecture that provably exclude periodic orbits whose length exceeds certain bounds instead of showing that very long orbits are rare in typical networks with certain architectures.  Note that bi-quadratic cooperative networks embody all the restrictions studied in the cited empirical work: The number of inputs of each regulatory function is restricted to two, each regulatory function is (nested) canalyzing, has a strong bias unless it is monic, and negative feedback loops are absent.  Nevertheless, as Part~I of this preprint \cite{PartI} shows, the combination of these properties still does not eliminate the theoretical possibility of existence of periodic orbits of length $> c^n$ for any $c < 2$.

The novelty of the results presented here is that if, in addition, all or a specified fraction of the regulatory functions are required to be non-monic, then there exist constants $c<2$ such that the resulting Boolean networks cannot have any periodic orbits of length $> c^n$ whatsoever.
As Theorem~\ref{GenTthm} shows, this will be true as long as there are fixed bounds on the number of inputs and outputs of each regulatory function as well as on their biases, regardless of the prevalence of negative feedback loops or whether the regulatory functions are actually canalyzing.

One can interpret Theorem~\ref{Turingthm} in a different way.
As noted earlier, variables with monic regulatory functions in cooperative Boolean systems just record the values of other variables at some time in the past (less than $n$ steps earlier).  Thus if we allow time delays in the definitions of regulatory functions, we can remove all but the
first variable on each `tape' and define a \emph{Boolean delay system} on the remaining variables that will have equivalent dynamics, in particular, that will have periodic orbits of the same length as the original system.

Formally, let $d$ be a positive integer and consider the state space $\Pi$ of an $M$-dimensional Boolean system and a map
$h$ so that  $h: \Pi^{[d]}  \rightarrow \Pi$.  Then $(\Pi, h)$ defines an $M$-dimensional \emph{Boolean delay system with
maximum delay $d$} whose
dynamics is given, for all times $t \geq d$ by

\begin{equation}\label{delay}
s(t+1) = h(s(t),s(t-1),...,s(t-d+1)).
\end{equation}

We associate to $h$ a function $H: \Pi^{[d]} \rightarrow \Pi^{[d]}$ defined by $H(s) = [s^2, \ldots , s^{d},
h(s)]$ for $s = [s^1, \ldots , s^d]$.
We will say that an $n$-dimensional Boolean system $(\Sigma, g)$ is \emph{induced by an $M$-dimensional Boolean delay system $(\Pi, h)$ with maximum delay $d$} if
there exists an injection $F = (F1, F2): [n] \rightarrow [M] \times [d]$ such that for the map  $f^*: \Pi^{[d]} \rightarrow \Sigma$ defined by
$(f^*(s))_i = s^{F2(i)}_{F1(i)}$ and any $f: \Sigma \rightarrow \Pi^{[d]}$ with $f^*\circ f = id$ we have

\begin{equation}\label{induced}
g(v) = f^*\circ H \circ f(v).
\end{equation}

In particular, as we indicated above, a cooperative $n$-dimensional Boolean system $(\Sigma, g)$ will be induced by a Boolean delay system with maximum delay $n$ whose
variables are only the ones with non-monic regulatory functions $f_k$, together with one variable from each `read-only' tape.
This allows us to reformulate Theorem~\ref{Turingthm} as follows:

\begin{theorem}\label{Delaythm}
Let $\alpha > 0$ and let $b$ be a positive integer. Then there exists a positive constant  $c < 2$ such that for sufficiently large $n$, every $n$-dimensional $b$-quadratic cooperative Boolean system with a periodic orbit of length at least $c^n$ is induced by a Boolean delay system with maximum delay at most $n$ and dimension at most
$\alpha n$.
\end{theorem}

It follows that all cooperative bi-quadratic Boolean systems with very long periodic orbits must be induced by Boolean delay systems with the same properties but of much smaller dimension.

Boolean delay systems as defined
 above are a special case of continuous-time Boolean delay systems as introduced in \cite{GhilI, GhilII}.
See \cite{GhilIII} for a comprehensive survey and additional references.  They may be relevant as a modeling tool for gene regulatory networks, since gene regulation always involves a delay between gene transcription and the time when the translated
gene product becomes available as a regulator, such as a transcription factor.  In fact, they are closely related to the conceptual framework of `kinetic logic' that was developed by R. Thomas in \cite{ThomasI, ThomasII} specifically for the study of gene regulation.

Our results appear to us of some philosophical interest given the metaphors discussed here and in the introduction.  Our theorems have the advantage
over numerical results of giving bounds that are rigorous and universally valid.  We want to emphasize though that our proof of Theorem~\ref{GenTthm} only shows existence and  gives an upper bound for $c(\eps, \alpha, b, r)$; we do not know whether this estimate is anywhere close to optimal.

For example,

$$c(0.25, 1, 2, 2) \leq 1.9959,$$

$$c(0.25, 0.1, 2, 2) \leq 1.9996.$$

Note that if a bi-quadratic cooperative system of sufficiently large dimension $n$ has a periodic orbit of length $> c(0.25, 0.1, 2, 2)^n$, then
at least $90\%$ of all regulatory functions must be monic; if such a system has a periodic orbit of length $> c(0.25, 1, 2, 2)^n$, then
at least some of the regulatory functions must be monic. On the other hand, Example~\ref{c=sqrt2} gives a lower bound

$$\sqrt{2} \leq c(0.25, 1, 2, 2).$$

It will be an interesting direction for future research to narrow the gap between the upper and lower bounds.

\section*{Acknowledgement}

We thank Xiaoping A. Shen for valuable suggestions on how to simplify our numerical estimates.


\begin{thebibliography}{2}


\bibitem{Chernoff} Chernoff, H. (1952). A measure of asymptotic efficiency for tests of a hypothesis based on the sum of observations.
 \emph{Ann. Math. Stat.} \textbf{23}, pp. 493-–507.

\bibitem{GhilI} Dee, D. and M. Ghil (1984). Boolean difference equations, I: Formulation and dynamic behavior. \emph{SIAM J. Appl. Math.}
\textbf{44}, 111--126.

\bibitem{DerridaStauffer} Derrida, B. and D. Stauffer (1986).
Phase Transitions in Two-Dimensional Kauffman Cellular Automata.
\emph{Europhys Lett.} \textbf{2}(10) 739--745.

\bibitem{PartI}  Enciso, G. A. and W. Just (2007). Large attractors in cooperative bi-quadratic Boolean networks. Part I.
\emph{Preprint.}
arXiv:0711.2799v2

\bibitem{GhilII} Ghil, M. and A. P. Mullhaupt (1985). Boolean delay equations. II: Periodic and aperiodic solutions.
\emph{J. Stat. Phys.} \textbf{41}, 125--173.

\bibitem{GhilIII} Ghil, M., Zaliapin, I. and B. Coluzzi (in press). Boolean Delay Equations: A Simple Way of Looking at Complex Systems.
arXiv:nlin/0612047v2.  To appear in \emph{Physica D.}

\bibitem{Harris} Harris, S. E., Sawhill, B. K., Wuensche, A. and S.
Kauffman (2002). A model of transcriptional regulatory networks based
on biases in the observed regulation rules. \emph{Complexity}
\textbf{7}(4),  23--40.

\bibitem{Hoeffding} Hoeffding, W. (1963).
Probability Inequalities for Sums of Bounded Random Variables.
\emph{J. Am. Stat. Assoc.} \textbf{58}(301), 13--30.

\bibitem{origins}  Kauffman, S. A. (1993) \emph{Origins of Order: Self-Organization and Selection in Evolution.}
Oxford U Press, 1993.

\bibitem{nestcan} Kauffman, S., Peterson, C., Samuelsson,
B. and C. Troein (2003). Random Boolean network models and the
yeast transcriptional network. \emph{PNAS} \textbf{100}(25),
14796--14799.

\bibitem{Landau} Landau, E. (1974). \emph{Handbuch der Lehre von der Verteilung der Primzahlen,} Chelsea Publishing Company, New York.

\bibitem{Okamoto} Okamoto, M. (1958). Some inequalities relating to the partial sum of binomial probabilities.
\emph{Annals of the  Institute of Statistical Mathematics} \textbf{10}, 29--35.

\bibitem{Sontag:Laubenbacher} Sontag, E., Veliz-Cuba, A., Laubenbacher, R. and A.S. Jarrah (in press).
The effect of negative feedback loops on the dynamics of Boolean networks. 
	arXiv:0707.3468v2 [q-bio.QM] To appear in \emph{Biophysical Journal.}

\bibitem{ThomasI} Thomas, R. (1973). Boolean formalization of genetic control circuits.
\emph{J. Theor. Biol.} \textbf{42}, 563--585.

\bibitem{ThomasII} Thomas, R. (1978). Logical analysis of systems comprising feedback loops.
\emph{J. Theor. Biol.} \textbf{73}, 631--656.

\bibitem{WeisbuchStauffer} Weisbuch, G. and D. Stauffer (1987).
 Phase-transition in cellular random Boolean nets,
\emph{J. de Physique} \textbf{48}(1), 11--18.

\end{thebibliography}
\end{document}